\documentclass[preprint,12pt]{elsarticle}

\usepackage{epsfig}

\usepackage{amsmath, amsthm, amscd, amsfonts, amssymb, graphicx, color}
\usepackage{float}
\usepackage{graphicx}
\usepackage{epstopdf}
\usepackage{subfigure}
\graphicspath{{fig/}}

\newtheorem{definition}{Definition}[section]
 
 \newtheorem{lemma}{Lemma}[section]

 \newtheorem{example}{Example}[section]
\journal{}

\begin{document}

\begin{frontmatter}


%\title{Title\tnoteref{label1}}
 %\tnotetext[label1]{}
 %\author{Name\corref{cor1}\fnref{label2}}
 %\ead{email address}
 %\fntext[label2]{}
 %\cortext[cor1]{}
%\address{Address\fnref{label3}}
 %\fntext[label3]{}

\title{A Novel Exploration of Diffusion Process based on Multi-types Galton-Watson Forests}


\author{Yanjiao Zhu$^1$}\author{Qilin Li$^2$}\author{Wanquan Liu$^{3*}$}\author{Chuancun Yin$^1$}\author{Zhenlong Gao$^1$}
 \address{$^{1}$School of Statistics, Qufu Normal University, Qufu, China}
 \address{$^{2}$ Department of Computing, Curtin University, Perth, Australia}
 \address{$^{3}$ School of Intelligent Systems Engineering, Sun Yat-sen University, Guangzhou, China\\ $^3$Email: liuwq63@mail.sysu.edu.cn}

\begin{abstract}
Diffusion is a commonly used technique for spreading information from point to point on a graph. The rationale behind diffusion is not clear. And the multi-types Galton-Watson forest is a random model of population growth without space or any other resource constraints. In this paper, we use the degenerated multi-types Galton-Watson forest (MGWF) to interpret the diffusion process and establish an equivalent relationship between them. With the two-phase setting of the MGWF, one can interpret the diffusion process and the Google PageRank system explicitly. It also improves the convergence behavior of the iterative diffusion process and Google PageRank system. We validate the proposal by experiment while providing new research directions.
\end{abstract}


\end{frontmatter}

% \linenumbers

\section{Introduction}
\label{}

The diffusion process is a phenomenon of spontaneous spreading of particles from a region of high concentration to one of the lower concentration regions. In other words, the diffusion process can be viewed as dynamic evolving when a system is not in equilibrium and random motion tends to bring its states towards uniformity. The diffusion processes have been used to model many physical, biological, engineering, economic, and social phenomena. Also, diffusion processes have been used in machine learning, because of its capacity in capturing the intrinsic manifold structure of data, and they have been widely used in many fields such as information retrieval \cite{YK,BY,EK,WT,DB}; clustering \cite{LL,DM}; saliency detection \cite{CZ,LM}; image segmentation \cite{WT,YP}; semi-supervised learning \cite{DB,WT2013}.

\begin{figure*}[htb]
   \centering
   \includegraphics[width=1.0\textwidth]{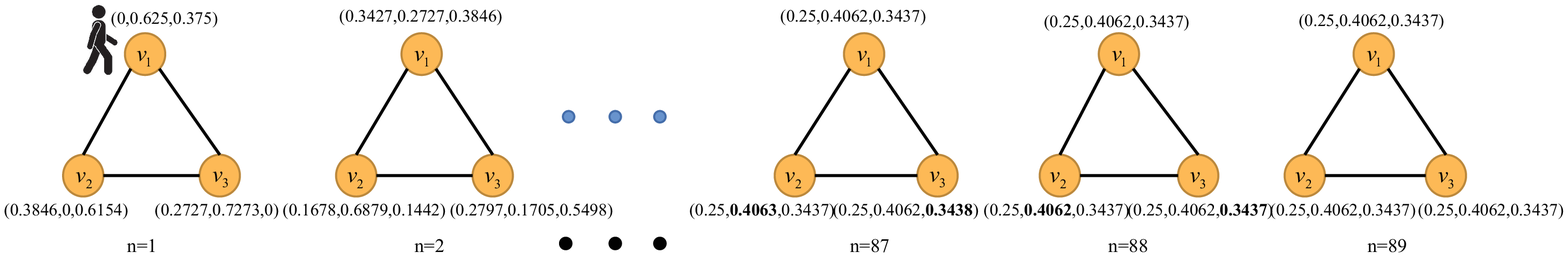}
   \caption{Random walk on the graph which contains three nodes. We take the accuracy of $1e^{-4}$ and find that after $88$ iterations $F_{n}$ reaches a stable state.}
   \label{randomwalk}
\end{figure*}

Technically, diffusion processes generally start from a given $N\times N$ affinity matrix $A$, which relates $N$ different elements to each other. An undirected graph $\mathcal{G}=(V, E)$ can be represented by an affinity matrix $A\in R^{N\times N},$ where the graph $\mathcal{G}=(V, E)$ consists of $N$ nodes $v_{i}\in V,$ and edges $e_{ij}\in E,$ that link nodes to each other. The edges between node $i$ and node $j$ can be weighted by their affinity value $A_{ij}.$ The diffusion process can be interpreted as a Markov random walk on graph $\mathcal{G}$, where the probability of walking from one node to another is governed by a transition matrix $P$, in which the relation between $A$ and $P$ is described as below. We first define a $N\times N$ diagonal matrix $D$ with $D_{ii}=\Sigma_{k=1}^{N}A_{ik},$ then the transition matrix $P$ can be written as follow:
\begin{eqnarray*}
P=D^{-1}A.
\end{eqnarray*}
Note that $P$ is a row-stochastic matrix (rows sum up to 1). The diffusion process is to place a particle on the node of an undirected graph $\mathcal{G}$, with the initial distribution $F_{0}$ which is an $N\times N$ matrix and contains the initial information for every node. In the existing literature on the diffusion process, $F_{0}$ was taken differently as $A$ (affinity matrix) \cite{JW}, $I$ (identity matrix) \cite{Zhu}, $P$ (transition matrix) \cite{CL}. In this paper, we let $F_{0}=P$ unless otherwise specified. Each particle transfers the information according to the transition matrix $P$, then the distribution of the particles that we get from one iteration is $F_{1}=F_{0}P$. Going through this process iteratively, the distribution of the particles will reach a stable state $\Pi$ which is the matrix of the left eigenvector of the transition matrix $P$ with corresponding eigenvalue $1$. This stable state is also the intrinsic manifold structure description of data. The random walk on the graph which contains three nodes is shown in Figure \ref{randomwalk}.

One important application for the diffusion process was the web PageRank \cite{BP}, in which each website corresponds to one node in the graph; the hyperlink between websites is represented by the edge in the graph. However, one typical problem is that the user sometimes browses the website by entering the URL rather than clicking on the hyperlink, which causes some influence on the PageRank system as prior information. The Google PageRank system \cite{BP} solved the problem by using the following algorithm:
\begin{eqnarray*}
F_{n}=\alpha F_{n-1}P+(1-\alpha)Y,
\end{eqnarray*}
where $Y$ represents the prior information of the web PageRank system. The Google PageRank system is an extension of the diffusion process.

In the literature, the diffusion process and its extensions were explained by the random walk model on the graph, but it has certain limitations. (1) We didn't have a clear understanding of how each particle moves, and its engineering significance was not obvious; (2) The transition matrix $P$ for the diffusion processes depends on the affinity matrix $A$ of the graph, which limits the expansion of the diffusion process; (3) For the Google PageRank system, the matrix $Y$ cannot be visually explained in the graph. To address these issues, this paper uses a special multi-type Galton-Watson forest (MGWF) model \cite{OG,MG} to explain the diffusion process from a different perspective.

\begin{figure}[H]
\centering
\subfigure[The 3-type GWF $\{\mathbf{Z_n}\}$. In the 1st generation the number of $1,2,3$-type are 1,1,1,respectively.]
{\begin{minipage}[t]{0.6\linewidth}
\centering
\includegraphics[width=3in]{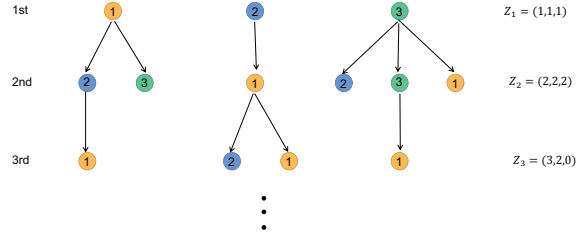}
%\caption{The 3-type GWF $\{\mathbf{Z_n}\}$. In the 1st generation the number of $1,2,3$-type are 1,1,1,respectively.}
\label{3mgwf}
\end{minipage}%
}\quad
\subfigure[MGWF with degenerated branching law. A parent particle can only produce one offspring particle.]{
\begin{minipage}[t]{0.6\linewidth}
\centering
\includegraphics[width=2in]{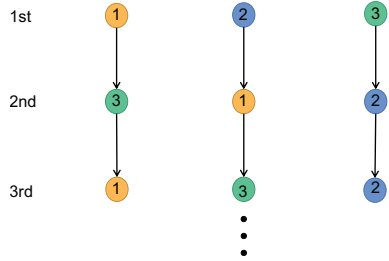}
%\caption{MGWF with degenerated branching law. A parent particle can only produce one offspring particle.}
\label{MGWFwithdegeneration}
\end{minipage}%
}%

%\centering
\caption{ The 3-type GWF and 3-type GWF with degenerated branching law}
\label{3generate}
\end{figure}

The MGWF is a generalization of mono-type Galton-Watson forests that describe the genealogy of a population in which individuals are differentiated by types. The MGWF $\{\mathbf{Z_n}\}$ has a visual representation of a multi-type branching process with multiple ancestors, where $\mathbf{Z_n}=\big(Z_{1},Z_{2},\cdots,Z_{N}\big)$, $Z_1$ represents the number of the particle of $1$-type,$\cdots$, and $Z_N$ represents the number of the particle for  the $N$-type, respectively. For a $i$-type parent, it lives  for a unit of time until upon death,  produces $j_1$ children of $1$-type, $\cdots$, $j_N$ children of $N$-type according to the branching law $P^{i}\big(j_{1},j_{2},\cdots j_{N}\big)$. As a counting process, MGWF mainly considers the ultimate property of population size $\{\mathbf{Z_n}\}$. An example of 3-type GWF is shown in Figure \ref{3mgwf}.

An MGWF with a degenerated branching law is not the focus of the research in the field of branching processes. However, it is of our interest in this paper to use it to explain the diffusion process. The degenerated branching law is: for every $i\in{1,2,\cdots,N.}$
\begin{eqnarray*}
\mathbf{P}^{i}\big(\{\mathbf{Z}\in\mathbb{N}^{N}:\Sigma_{j=1}^{N}Z_{j}\ne1\big\})=0
\end{eqnarray*}
 That is, a parent can produce only one offspring. Essentially, the members of this forest are infinite direct trees as shown in Figure \ref{MGWFwithdegeneration}. The N points in the random walk correspond to the N types of particles in the MGWF. The random walk on the graph is pulled longitudinally into particles to produce offspring by this way, in which the concept of "level" comes up.

\begin{figure}[H]
\centering
\includegraphics[width=0.5\textwidth]{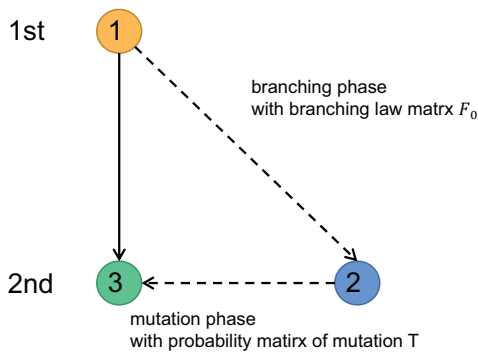}
\caption{MGWF with mutation. A unit of time is divided into two phases - branching phase and mutation phases.}
\label{2phases}
\end{figure}

 To use the MGWF to explain the diffusion process explicitly, we innovatively introduce the concept of particle mutation (in a unit of time, $i$-type particles produce $j$-type particles, and there is a chance that the $j$-type particle will mutate into $k$-type particles). Therefore, the random process can be divided into two phases - branching phase and mutation phase as shown in Figure \ref{2phases}. By dividing each step of the diffusion process (It corresponds to the unit time in the MGWF) into two phases, one not only can understand the diffusion process more clearly, but also it would inspire us to learn the diffusion process in a general setting.

With this two-phase setting, we can further consider MGWF with immigration \cite{S}. The classical theory of immigration is to introduce immigrants to ensure that the entire forest would not extinct so that immigrant particles enter the forest and produce offspring together with the original particles. However, we only consider immigration particles as another option different from the branching-mutation mechanism to produce offspring. The model of MGWF with immigration and mutation can be used to explain the well-known Google PageRank system, which is a successful extension of the diffusion process. This will also inspire us to interpret the Google PageRank system from a new perspective.

In summary, the main contributions of this article are as follows:

\begin{itemize}
\item[$\bullet$] By introducing the novel concept of mutation into the MGWF with the degenerated branching law, each step of the diffusion process is divided into two phases - the branching phase and the mutation phase. Such explanation provides novel insight on the rationale of the diffusion process from a longitudinal perspective, compared to the existing random walk interpretation in the literature.
\item[$\bullet$] In the existing diffusion process, the transfer matrix $P$ depends on the affinity matrix $A$. But by looking at the diffusion process from the perspective of the MGWF, the $P$ and $A$ can be separated as unrelated parts, which is a benefit for the extension of a more general diffusion process.
\item[$\bullet$] Within the framework of branching-and-mutation, we propose a variant of the diffusion process in which the mutation probability can change. It is a more general diffusion process that includes the original as a special case.
\item[$\bullet$] We further extend the MGWF by introducing immigration, which can interpret the Google PageRank system well.
\item[$\bullet$] Finally, two examples (Example \ref{changeT},\ref{changeY}) of the variant of diffusion process in which mutation probability can be changed prove that the variant of diffusion process can effectively improve the convergence rate of iteration.
\end{itemize}

This paper is organized as follows. In the preliminary section, we introduce the MGWF model in detail; In Section \ref{Mutations}, we introduce the concept of mutation into the MGWF with degenerated branching law and explain the diffusion process under our new model; also, We introduce the MGWF model with immigration and use it to explain the Google PageRank system in section \ref{Google}; in section \ref{Simulations}, we provide some examples of the diffusion process and the Google PageRank system respectively, and show that possible modifications in the setting can increase the rate of convergence in the extended diffusion process; We draw conclusions and point out potential future directions in section \ref{Conclusions}.

\section{Preliminaries}
\label{Preliminaries}
\subsection{Random walk and Branching process}
\label{}
First, let us look at the definition of the Markov chain.
\begin{definition}$\mathbf{Markov~chain}$
Let $\{X_{n},n\ge 0\}$ be a sequence of discrete random variables. For states space $\mathbb{S}$, the $\{X_{n},n\ge 0\}$ satisfying the Markov properties:
\begin{eqnarray*}
\mathbf{P}(X_{n+1}=j|X_{n}=i,\cdots ,X_{0}=i_{0})=\mathbf{P}(X_{n+1}=j|X_{n}=i)=\mathbf{P}_{i,j}
\end{eqnarray*}
for all $n=1,2,3\cdots$ and states $i_{0},\cdots,i,j\in\mathbb{S}$, is called a Markov chain with initial distribution $F_{0}$ and transition probability matrix $P$. A finite-state Markov chain is a Markov chain in which $\mathbb{S}$ is finite.
\end{definition}

A very important special case of the Markov chain is the random walk on an undirected graph. The random walk picks a neighbor at each step randomly and moves to that neighbor. Hence, the transition matrix is $P=D^{-1}A$, where $A$ is a $N\times N$ affinity matrix of the graph, and $D$ is the diagonal matrix with $D_{ii}=\Sigma_{j=1}^{N}A_{ij}$.

 The simplest type of a branching process is the Galton-Watson process (GW process) which is also a Markov chain. The branching process is one of the classical fields in applied probability and it can deal with a mathematical representation for the population changes in which their members can reproduce and die, subject to laws of chance. The particles may be of different types depending on their age, energy, position, or other factors. However, the type of all particles is the same in the GW process. Let us define a GW process as a Markov chain $\{Z_n\} =Z_{0},Z_{1},Z_{2},\cdots$ , where $Z_n$ is a random variable describing the population size at the $n$-th generation. The Markov property can be interpreted as: in the $n$-th generation, the $Z_{n}$ individual independently gives rise to some number of offsprings $\xi_{n+1,1}, \xi_{n+1,2}, \cdots, \xi_{n+1,Z_{n}}$ for the $(n+1)$-th generation. $\xi_{n+1,j}$ can be thought of as the number of members in the $(n+1)$-th generation which are offsprings of the $j$-th member of the $n$ generation. Note that $\{\xi_{n,j} , n \ge 1, j \ge 1\}$ are identically distributed (having a common distribution $\{p_{k}\})$ non-negative integer-valued random variables. Thus, the cumulative number produced for the $(n+1)$-th generation is
\begin{eqnarray*}
Z_{n+1} = \xi_{n+1,1} + \xi_{n+1,2} + \cdots + \xi_{n+1,Z_{n}}.
\end{eqnarray*}
Thus the probability of any future behavior of the process, when its current state is known exactly, can not be altered by any additional knowledge concerning its past behavior.

\begin{figure}[H]
\centering
\subfigure[The random walk on a graph with 5 vertices.]{
\begin{minipage}[t]{0.5\linewidth}
\centering
\includegraphics[width=2.5in]{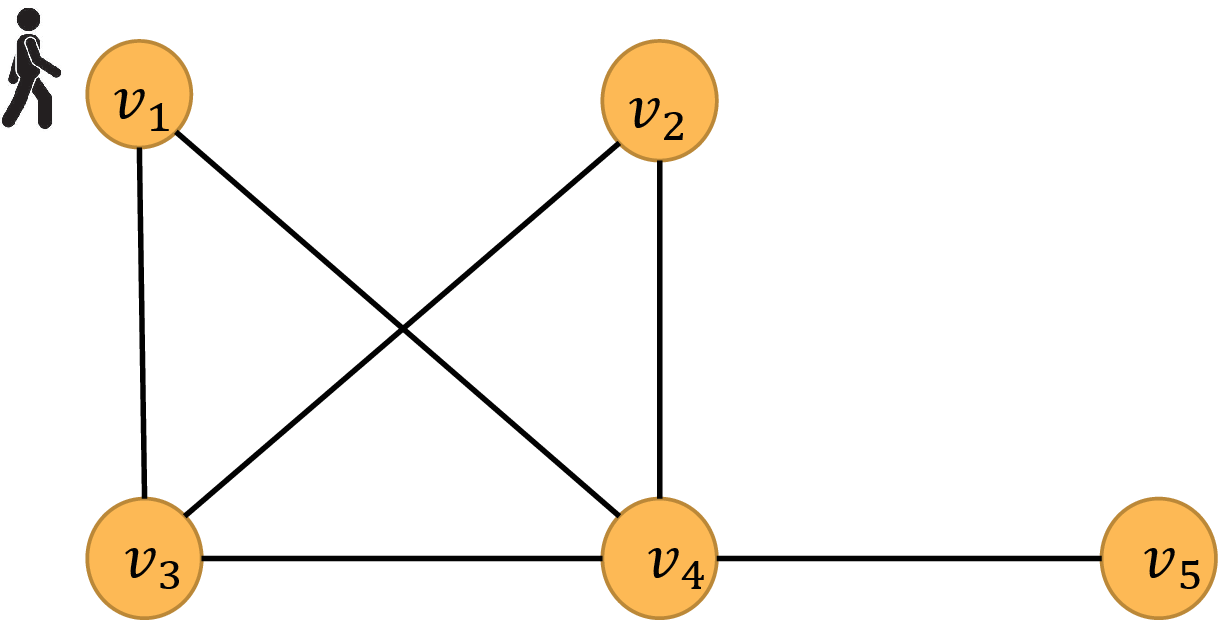}
%\caption{}
%\label{3mgwf}
\end{minipage}}%
\subfigure[The Galton-Watson process.]{
\begin{minipage}[t]{0.5\linewidth}
\centering
\includegraphics[width=2.5in]{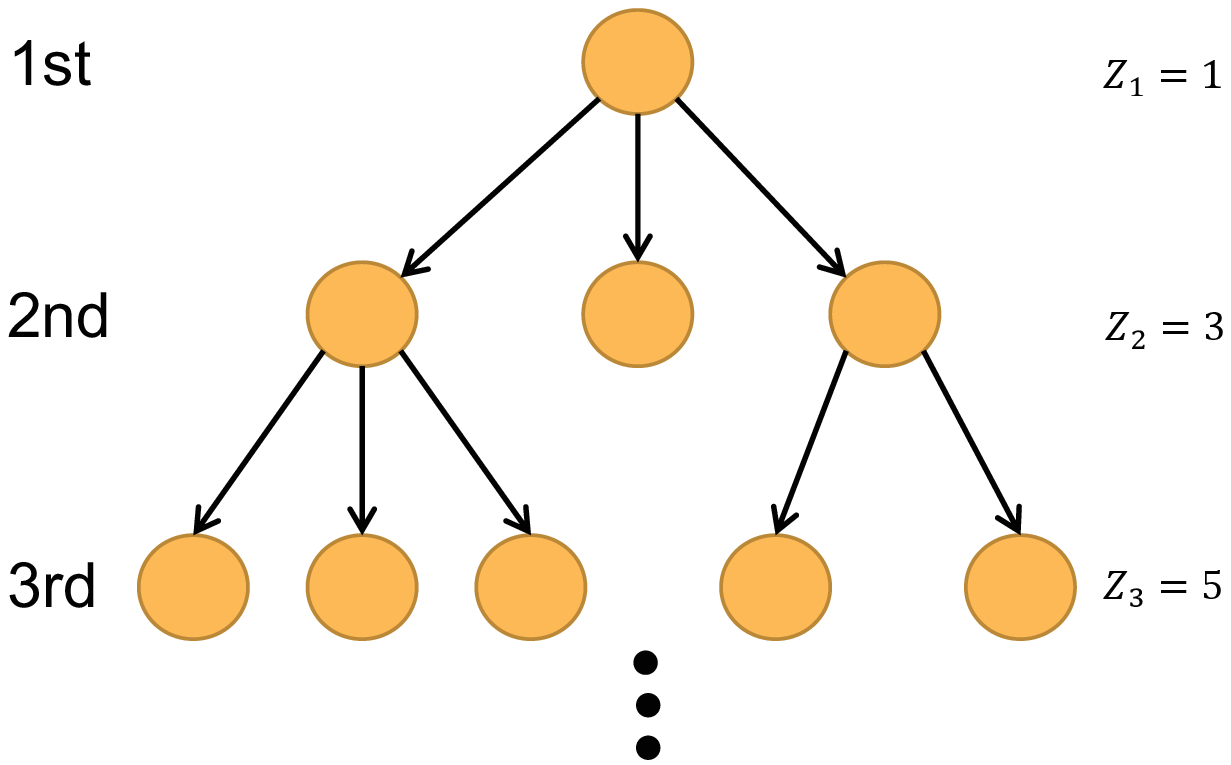}
%\caption{MGWF with degenerated branching law. A parent particle can only produce one offspring particle.}
%\label{MGWFwithdegeneration}
\end{minipage}}%
%\centering
\caption{ The random walk and the GW process}
\label{RB}
\end{figure}

Two examples of Markov chains are shown in Figure \ref{RB}; one as random walks on an undirected graph and another as a GW process separately.
\subsection{The multi-type Galton-Watson forest with degenerated branching law}

In this section, we consider a multi-type GW process without the assumption that all particles are of the same type. Set a N-type GW process $\{\mathbf{Z}_{n}\}_{n\ge 0}$, with the initiated state as
\begin{eqnarray*}
\mathbf{Z}_{0}=\mathbf{i}=(i_{1},\cdots,i_{N})
\end{eqnarray*}
where $i_{k}\in\mathbb{Z}_{+}=\{0,1,2,\cdots\},k=1,\cdots N$, $i_{k}$ is the number of individuals of $k$-type in the first generation. For the process with $\mathbf{Z}_{0}=\mathbf{i}$, the number of $j$-type individuals in the $n$-th generation is
\begin{eqnarray*}
\mathbf{Z}_{n}^{(\mathbf{i})}=(Z_{n,1}^{(\mathbf{i})},Z_{n,2}^{(\mathbf{i})}\cdots,Z_{n,N}^{(\mathbf{i})})
\end{eqnarray*}
where $Z_{n,j}^{(\mathbf{i})}$ represents the number of particles of $j$-type  in the $n$-th generation.  We assume that the parent $\mathbf{i}$, lives for a unit of time and, upon death, produces children of all types and according to the offspring distribution
\begin{eqnarray*}
\{P^{\mathbf{i}}(\mathbf{j})\}=\{P^{(i_{1},i_{2},\cdots,i_{N})}\big(j_{1},j_{2},\cdots,j_{N}\big)\},i_{k},j_{k}\in \mathbb{N}, k=1,2,\cdots,N
\end{eqnarray*}
and the independence of other individuals. In which
$\{P^{(i_{1},i_{2},\cdots,i_{N})}\big(j_{1},j_{2},\cdots,j_{N}\big)\}$ is the probability that the parents produce $j_1$ children of $1$-type, $\cdots$ , $j_N$ children of $N$-type, and the parents are $i_1$ particles of $1$-type, $\cdots$ , $i_N$ particles of $N$-type. The case with the initial number of particles being  multiple is vividly called a multi-type Galton-Watson forest (MGWF).

Let $\mathbf{e}_{i},i=1,2,\cdots ,N$ denote a N-dimensional row-vector with $i$th component as $1$, and the remaining being zero.

Next, a special N-type Galton-Watson forest is considered. There are $N$ particles in the system at the time $0$, and their types are $1,2,\cdots,N$ respectively. Each particle is considered as an ancestor of a multi-type GW process, i.e.
\begin{eqnarray*}
\mathbf{Z}_{0}=\mathbf{e}_{i}
\end{eqnarray*}
We denote the initial state of the MGWF as an $i$-type particle. Each ancestor lives for a unit of time and may produce children of all types upon death according to the offspring distribution $\{P^{\mathbf{e}_i}(\mathbf{e}_{j})\}$. For convenience, we abbreviate it as:
\begin{eqnarray*}
\{P^{\mathbf{e}_i}(\mathbf{e}_{j})\}=\{P^{i}(j)\},i,j=1,2,\cdots,N,
\end{eqnarray*}
where $\Sigma_{j=1}^{N} P^{i}(j)=1$, in other words, for every $i\in{1,2,\cdots,N}$
\begin{eqnarray*}
\mathbf{P}^{i}\big(\{\mathbf{Z}\in\mathbb{N}^{N}:\Sigma_{j=1}^{N}Z_{j}\ne1\big\})=0
\end{eqnarray*}
i.e., each parent produces only one offspring. In this way, the number of multi-type branching tree formed in the forest is $N$. From \cite{OG,MG}, we know the system is the MGWF with degenerate branching law, i.e. every element in the forest is an infinite trees. However, here we still call this system a MGWF in this paper. The system with a $N\times N$ branching law matrix $P$
\begin{eqnarray*}
P=\big(P^{i}(j)\big)_{i,j=1,2,\cdots,N}
\end{eqnarray*}
where for any $i=1,2,\cdots,N$, we have $\Sigma_{j=1}^{N} P^{i}(j)=1$. The difference between the MGWF and the MGWF with degenerated branching law is graphically illustrated in Figure \ref{3generate}.

In the existing literature \cite{OG,MG,DL,D,DG}, many MGWFs mainly discuss the convergence of their corresponding production process, which is the sequence of generations of the individuals of the MGWF visited in lexicographical order. It is easy to check that the height process fully characterizes the forest. In the literature, it is assumed that the branching law is immutable. In this paper, we mainly discuss the change of the branching law in some special cases.

\begin{example}
Let us first consider an example where there are three types of particles in the system, i.e. $N=3$. The 3-type GW forest with the degenerated branching law $P$ is shown in Figure \ref{MGWFwithdegeneration}. The branching law matrix is
\begin{eqnarray*}
P_{3\times 3}=\left(\begin{matrix}
0& 0.6250&0.3750\\
0.3846&0&0.6154\\
0.2727&0.7273&0
\end{matrix}
\right).
\end{eqnarray*}
\end{example}

\section{Mutations in the MGWF}
\label{Mutations}
\subsection{The Mutations}
Let us consider mutation in the framework of MGWF with degenerated branching law.``Mutation" is not a new concept in the literature of the MGWF \cite{CN,DT}. The original definition is that a mutation producing $i$-type is the birth event of an individual of $j$-type from an individual of any type $j\ne i$. In this paper, we propose a novel variant of mutation which can better conform to the actual situation, and more importantly, it provides us more insight into the rationale of the diffusion process by separating it into two phases: branching and mutation.

\begin{figure}[H]
\centering
\includegraphics[width=0.6\textwidth]{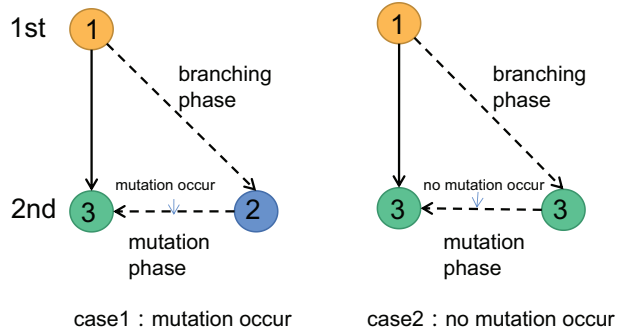}
\caption{The two cases in which mutations occur(case 1) and do not occur(case 2) during the mutation phase  }
\label{F01}
\end{figure}

\begin{definition}{$\mathbf{(Mutation)}$}
When mutation time is ignored, a unit of time is divided into two phases- the branching phase and the mutation phase.

(i) In the branching phase, an $i$-type particle produces a $j$-type particle according to the branching law $P^{i}(j)$.

(ii) In the mutation phase. The mutation occurs:  the $j$-type particle may mutate into a $k$-type particle according to the probability of mutation $T^{j}(k),j\neq k,i,j,k=1,2,\cdots,N$;

No mutation occurs: $j$-type particle may also remain type unchanged with probability $T^{j}(j)$. The two cases in which mutations occur and do not occur during the mutation phase are shown in Figure \ref{F01}.
\end{definition}

Naturally, the branching law $\{P^{i}(j)\}_{i,j=1,\cdots,N}$ of each generation remains the same. However, when mutations are considered, the branching law $\{P^{i}(j)\}_{i,j=1,\cdots,N}$ of each generation will be different, i.e., the branching law matrix will change, and the iterative process is as follows.

Suppose that at time $0$, there is an $i$-type ancestor in the system, which is the $1st$ generation. After a unit of time, we observe that the $i$-type ancestor has a $j$-type offspring. Then in the $2nd$ generation, the probability of an $i$-type particle produces a $j$-type particle when mutations are taken into account:
\begin{eqnarray}
\label{P2}
P^{i}_{2}(j)=P_{1}^{i}(j)T^{j}(j)+\Sigma_{k\neq j}P_{1}^{i}(k)T^{k}(j)
\end{eqnarray}
\begin{eqnarray*}
k=1,2,\cdots,N. and ~k\neq j, P^{i}_{1}(j)=P^{i}(j)
\end{eqnarray*}
where the subscript denotes the generation. The first term on the right-hand side of equation \ref{P2} is the probability of no mutation occurring in the mutation stage, and the second term on the right-hand side of equation (1) is the probability of the mutation occurring. Its matrix form is :
\begin{eqnarray*}
P_{2}=P_{1}T
\end{eqnarray*}
We call the matrix $T$ a probability matrix of mutation. Iteratively, for the $nth$ generation, we have
\begin{eqnarray*}
P_{n}=P_{n-1}T=PT^{n-1}
\end{eqnarray*}

Next, we use one example to show the iteration process.
\begin{example}{(The continuation of example 2.1)}
Here we consider the special case, the probability matrix of mutation $T=P$. The 3-type GW forest with mutation and the mutation details are shown in Figure \ref{eg1}.
\begin{figure}[H]
\centering
\subfigure[The 3-type GW forest with mutation.]{
\begin{minipage}[t]{0.5\linewidth}
\centering
\includegraphics[width=2.8in]{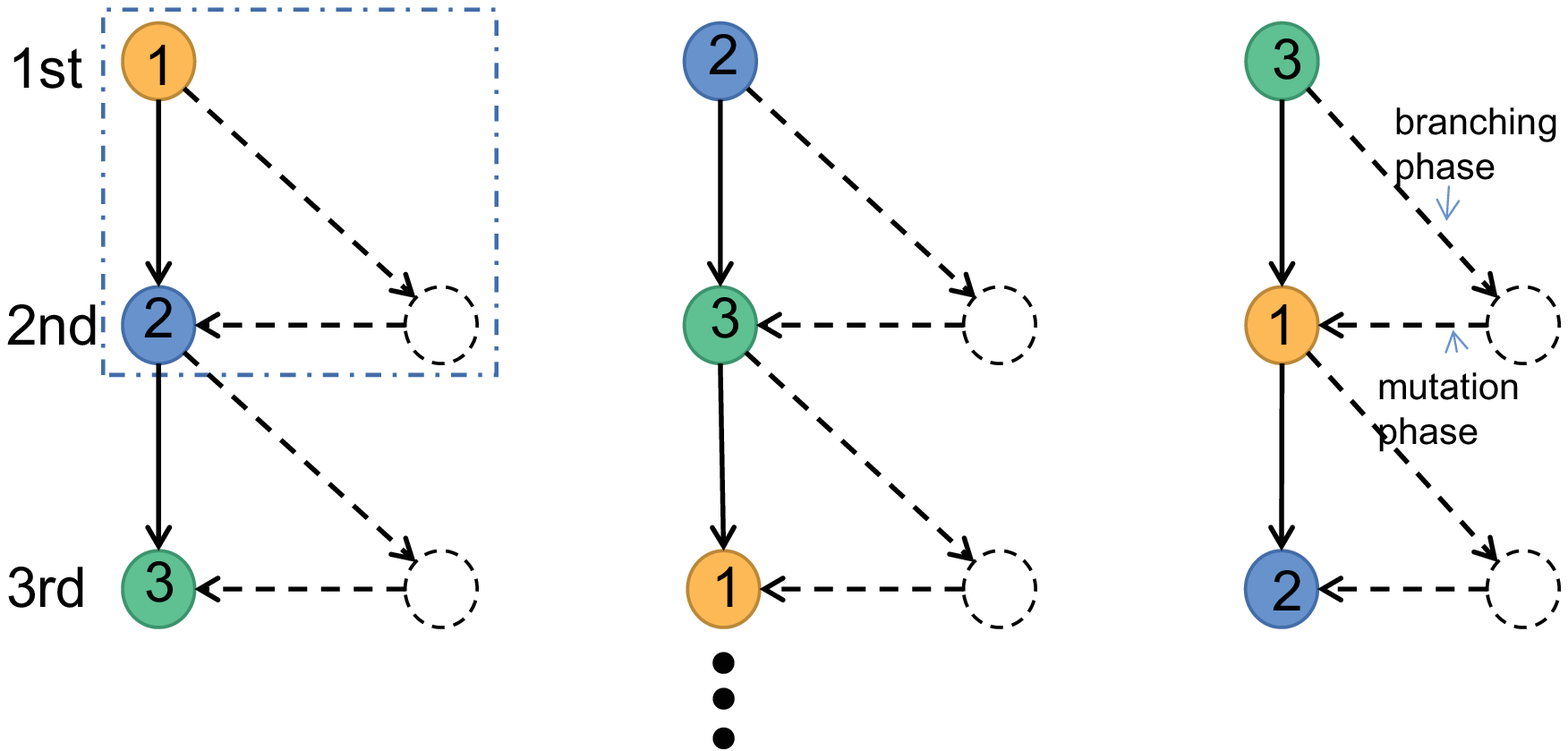}
%\caption{}
%\label{3mgwf}
\end{minipage}}%
\subfigure[The mutation details of the 3-type GW forest with mutation.]{
\begin{minipage}[t]{0.5\linewidth}
\centering
\includegraphics[width=2.5in]{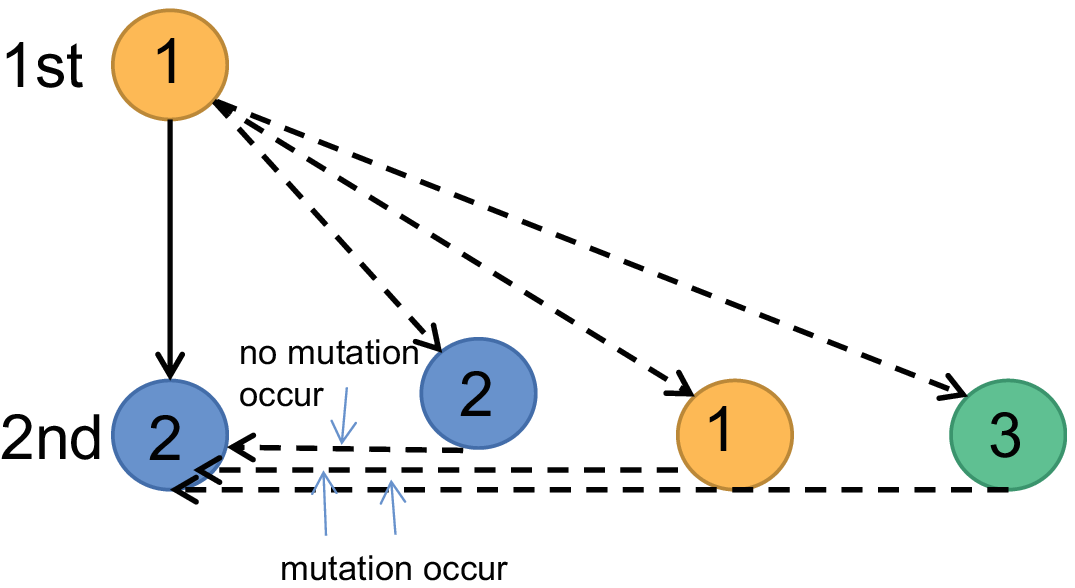}
%\caption{MGWF with degenerated branching law. A parent particle can only produce one offspring particle.}
%\label{MGWFwithdegeneration}
\end{minipage}}%
%\centering
\caption{ The 3-type GW forest with mutation and the mutation details }
\label{eg1}
\end{figure}
where
\begin{eqnarray*}
P_{2}&=&P*T\\
&=&\left(\begin{matrix}
0& 0.6250&0.3750\\
0.3846&0&0.6154\\
0.2727&0.7273&0
\end{matrix}
\right)*\left(\begin{matrix}
0& 0.6250&0.3750\\
0.3846&0&0.6154\\
0.2727&0.7273&0
\end{matrix}
\right)\\
&=&
\left(\begin{matrix}
0.3427& 0.2727&0.3846\\
0.1678&0.6879&0.1442\\
0.2797&0.1705&0.5498
\end{matrix}
\right)
\end{eqnarray*}
and
\begin{eqnarray*}
P_{3}=P_{2}*T=P*T^{2}=\left(\begin{matrix}
0.2098& 0.4939&0.2963\\
0.3039&0.2098&0.4863\\
0.2155&0.5747&0.2098
\end{matrix}
\right).
\end{eqnarray*}
\end{example}

Next, we will consider the convergence of the above iterative process and we have the following basic result for the convergence.

\subsection{The stable state}
Firstly, we introduce a few concepts and lemmas that come from literature \cite{LLM,OH}.
\begin{definition}{$\mathbf{(communicate)}$}
Let $i,j\in \mathbb{S}$ be two states of a Markov chain with transition matrix $T$. Then state $i$ communicates with state $j$ if there exist nonnegative integer $m$ and $n$ such that $T^{m}_{j,i}>0$ and $T^{n}_{i,j}>0$.
\end{definition}
\begin{definition}{$\mathbf{(irreducible)}$}
A Markov chain with finite-state space $\mathbb{S}$ and transition matrix $T$ is said to be irreducible if for all $i,j\in \mathbb{S}$ are communicated.
\end{definition}
\begin{definition}{$\mathbf{(aperiodic)}$}
The period $d(i)$ of a state $i$ of a Markov chain is the greatest common divisor of all time steps $n$ for which $T^{n}_{ii}>0$. If the period is $1$, the state is aperiodic.
\end{definition}
If a Markov chain is irreducible, the period of the chain is the period of its single communication class. If the period of every communication class is $d=1$, then the Markov chain is aperiodic.
\begin{definition}{$\mathbf{(regular)}$}
A stochastic matrix $T$ is regular if some power $T^k$ contains only strictly positive entries.
\end{definition}
\begin{lemma}
\label{regular}
Let $T$ be the transition matrix for an irreducible, aperiodic Markov chain. Then $T$ is a regular matrix.
\end{lemma}
\begin{lemma}
\label{convergence}
If $T$ is a regular $m\times m$ transition matrix of Markov chain with $m\ge 2$, then the following statements are all true.

(1) There is a stochastic matrix $\Pi$ such that $\lim_{n\to \infty }T^{n}=\Pi$.

(2) Each row of $\Pi$ is the same probability vector $\boldsymbol{\pi}$.

(3) For any initial probability vector $\mathbf{x_0},~\lim_{n\to \infty}\mathbf{x_0} T^n =\boldsymbol{\pi}$

(4) The vector $\boldsymbol{\pi}$ is the unique probability vector that is an eigenvector of $T$ associated with the eigenvalue $1$,i.e.$\boldsymbol{\pi}=\boldsymbol{\pi}T$.

\end{lemma}

As described above, to prove the convergence of the iterative process we make the following statement.
\begin{itemize}
\item[$\bullet$] The MGWF with mutation is a finite-state Markov chain. The state-space of MGWF with mutation $\mathbb{S}$ is the finite type space of particles $\{1,2,\cdots,N\}$. And the iterative process $F_{n}=F_{n-1}T$ implies that $F_{n}$ depends only on $F_{n-1}$ which is meets the Markov property.
\item[$\bullet$] The MGWF with mutation is an irreducible chain. Different types of particles can mutate with each other to ensure that all states of the MGWF with mutation $\{1,2,\cdots,N\}$ are communicating.
\item[$\bullet$] The MGWF with mutation is required to be an aperiodic chain.
\end{itemize}

  From Lemma \ref{regular}, the probability matrix of the mutation $T$ is regular. Therefore, from Lemma \ref{convergence}, we have that the  MGWF with the mutation has a unique stationary distribution $\Pi$, s.t.
\begin{eqnarray*}
\lim_{n\to \infty }T^{n}=\Pi
\end{eqnarray*}
where Each row of $\Pi$ is the same probability vector $\boldsymbol{\pi}$ which is the unique probability vector that is an eigenvector of $T$ associated with the eigenvalue $1$.

When the system of the MGWF with mutations reaches a stable state $\Pi$, the branching law is no longer affected by the mutation.

\begin{example}{The continuation of example 3.1}
 We still consider the 3-type GW forest with mutation. Where the branching law matrix is
\begin{eqnarray*}
P=\left(\begin{matrix}
0& 0.6250&0.3750\\
0.3846&0&0.6154\\
0.2727&0.7273&0
\end{matrix}
\right)
\end{eqnarray*} and the probability matrix of mutation $T=P$.
We verify that $k=2,T^{2}_{ij}>0$ for all $i,j\in \{1,2,\cdots,N\}$, so $T$ is regular, satisfying the Lemma \ref{convergence} condition.
We set the accuracy to $1e^{-4}$ and found that when $n=88$ the
\begin{eqnarray*}
P^{88}=\left(\begin{matrix}
0.2500& 0.4062&0.3437\\
0.2500&0.4062&0.3437\\
0.2500&0.4062&0.3437
\end{matrix}
\right)
\end{eqnarray*}
and iterating $P^{89},P^{90},\cdots$ again no longer changes, i.e., the particle mutation will not affect the branching law of the MGWF.
\end{example}

\subsection{Interpretation of diffusion process}

A diffusion process, in general, starts from a $N\times N$ affinity matrix $A$, which relates $N$ different elements to each other. We interpret the matrix $A$ as a graph $\mathcal{G}=(V,E)$, consisting of $N$ nodes $v_{i}\in V$, and edges $e_{ij}\in E$ that link nodes to each other, fixing the edge weights to provided affinity values with $A_{ij}$. Diffusion processes then spread the affinity values through the entire graph, based on the defined edge weights.
 When using a random walk on a graph to explain the diffusion process, there are three steps: (1) initialization; (2) definition of the transition matrix; (3) definition of the diffusion process.

 In this paper, we use the model of MGWF with mutation to explain the diffusion process from a new perspective. We can regard the $N$ data points as $N$ different particle types, the transition of the random walk between nodes corresponds to the iteration of MGWF with mutation. And the unit of time for each iteration is divided into two phases: the branching phase and the mutation phase.

 Assume the initial branching law and the probability matrix of mutation are given. To ensure that the initial branching law is a row-stochastic matrix, we take $F_{0}=P=D^{-1}A$ where $D=diag(D_{ii}),D_{ii}=\Sigma_{j=1}^{N}A_{ij}$, and especially the probability matrix of mutation $T=P=D^{-1}A$. After the first unit time, the branching law of the first generation is given by
 \begin{eqnarray*}F_{1}=F_{0}T.\end{eqnarray*}
 Since the mutation probability in the mutation phase is a constant, the branching law of the second generation is derived as  \begin{eqnarray*}F_{2}=F_{1}T=F_{0}T^{2},\end{eqnarray*}
 Eventually, The branching law of the $nth$ generation is \begin{eqnarray*}F_{n}=F_{n-1}T=F_{0}T^{n}.\end{eqnarray*}
 It can be seen from section 3.2 that the branching law will reach a stable state, that is, even if the mutation occurs, the branching law will no longer be affected by the mutation.

The correspondence between the diffusion process and the MGWF with the mutation is shown in Table \ref{diffusion and MGWF}.
\begin{table*}[th]
\begin{center}
\caption{The correspondence between diffusion process and MGWF with mutation }
\label{diffusion and MGWF}
\begin{tabular}{ccc}
\hline
Diffusion process&{}&MGWF with mutation\\
\hline
N nodes in the graph&{}&N types of particle\cr
Edges in the graph&{}& Particle type change\cr
An iteration of the diffusion process&{}&A unit time of MGWF\cr
Initialization $F_{0}=P$&{}&Initial branching law matrix $F_{0}=P$\cr
Transition matrix $P$&{}&The probability matrix of mutation $T=P$\cr
Update scheme $F_{n}=F_{n-1}P$&{}&Update scheme $F_{n}=F_{n-1}T$\cr
       \hline
\end{tabular}
\end{center}
\end{table*}

In this section, we established the relationship between the diffusion process and the MGWF with mutation and the convergence of the diffusion process is proved indirectly via the convergence of the MGWF. Next, we will propose the immigration concept for the MGWF and discuss its relationship with the Google Pagerank system, which is an extension of the diffusion process.

\section{The MGWF with immigration and the Google PageRank system}
\label{Google}
\subsection{The MGWF with mutation and immigration}

\begin{figure}[H]
\centering
\subfigure[The 3-type branching process with immigration.]{
\begin{minipage}[t]{0.5\linewidth}
\centering
\includegraphics[width=2.5in]{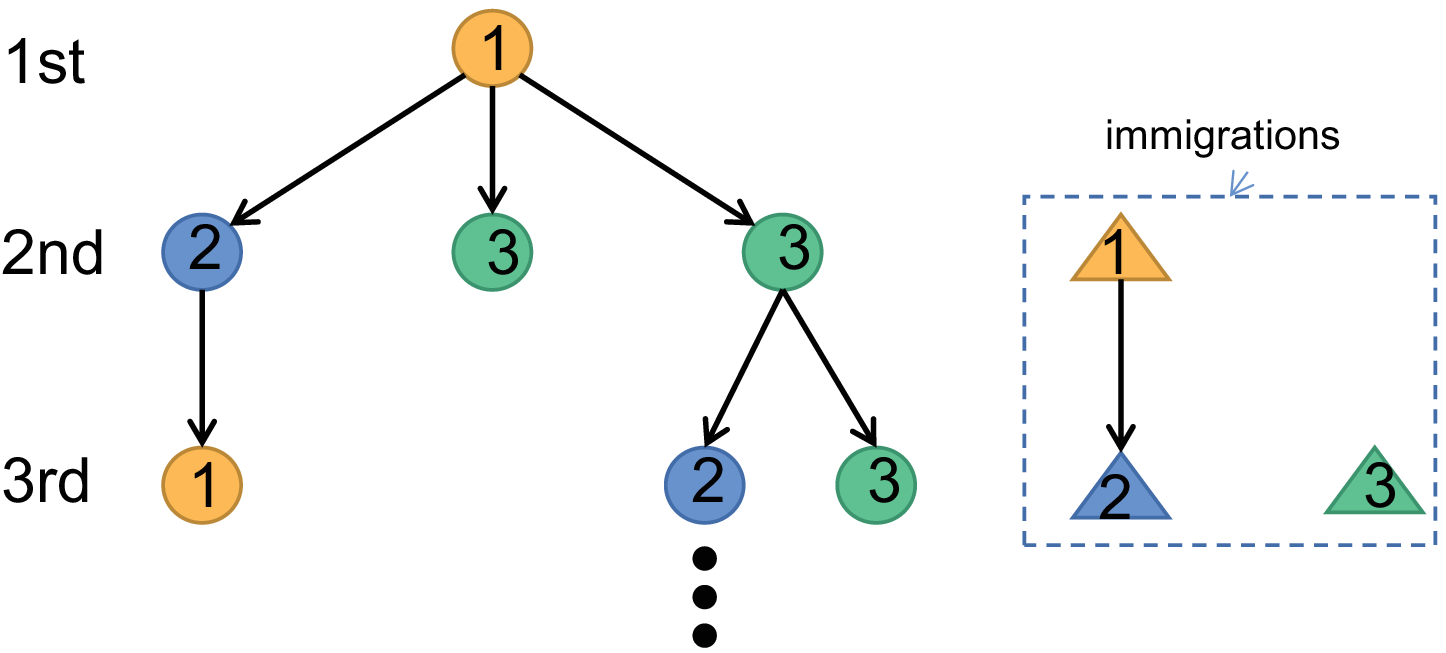}
%\caption{}
\label{Bimmigration}
\end{minipage}}%
\subfigure[The relationship between immigration mechanism and the branching-mutation mechanism in the MGWF.]{
\begin{minipage}[t]{0.5\linewidth}
\centering
\includegraphics[width=2.5in]{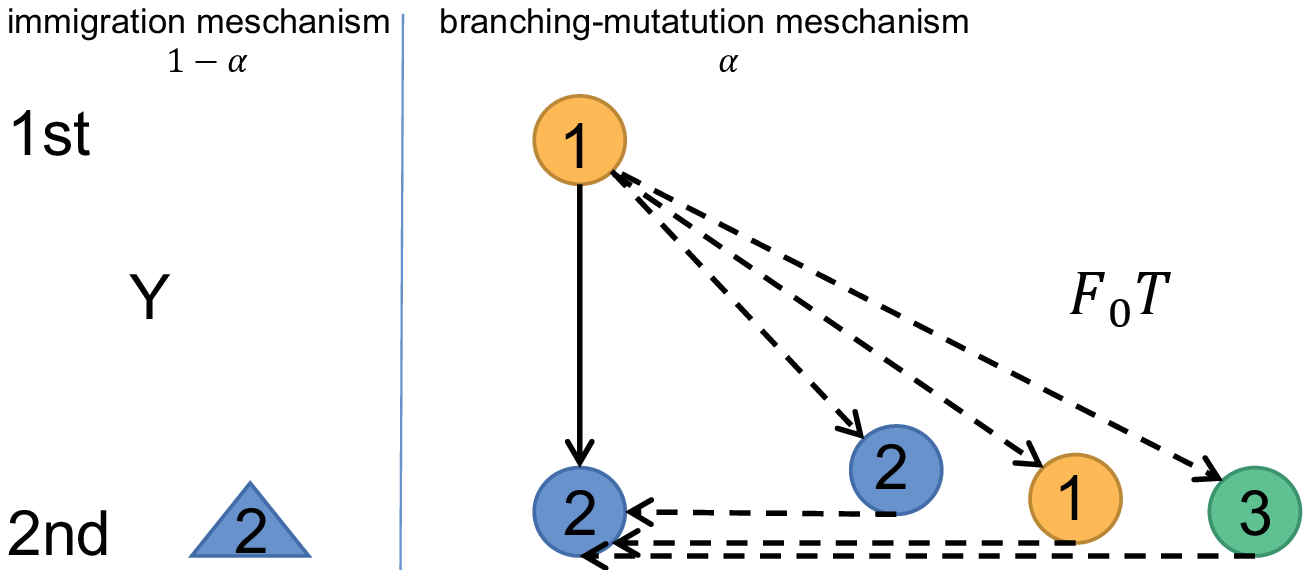}
%\caption{MGWF with degenerated branching law. A parent particle can only produce one offspring particle.}
\label{immigration}
\end{minipage}}%
%\centering
\caption{ The immigration in branching process and MGWF with mutation }
\label{eg}
\end{figure}

The classic MGWF has a wide range of extensions, and one of the important extensions is the GW process with immigration \cite{S,S1970,LZ,SZ}. The MGWF with immigration in the existing literature can be divided into states-dependent immigration and states-independent immigration. In this article, we consider states-independent immigration. The classical theory of immigration is to introduce immigrants to ensure that the entire forest is not extinct so that immigrant particles would enter the forest and can produce offspring together with the original particles, as shown in Figure \ref{Bimmigration}. In this section, we only consider immigration particles as another option different from the original branching to produce offspring in addition to the original branching. That is, the branching-mutation mechanism will no longer work if immigrants enter the forest.

In a $N-$type GWF system, for every tree and every generation, there maybe an immigrant of $j$-type particle entering with probability $b_{j}$, where $j=1,2,\cdots,N$ and $\Sigma_{j=1}^{N}b_{j}=1$. We allow migrants to enter a MGWF with mutations systems. When it is observed that an $i$-type particle in the $1$-st generation has a child particle of $j-$type in the second generation, there are two possible mechanisms for this phenomenon:

(1) Generated by a branching-mutation mechanism with probability $\alpha$. i.e. An $i$-type particle produces a $j$-type particle in consideration of mutations. Where $\alpha\in(0,1)$.

(2) Acted by the immigration mechanism with probability $(1-\alpha)$. i.e. in the second generation, an immigrant particle of $j$-type enters directly, the branching - mutations mechanism no longer works.

Therefore, the branching law in the second generation will change as follows:
\begin{eqnarray*}
P^{i}_{2}(j)=\alpha\big(P_{1}^{i}(j)T^{j}(j)+\Sigma_{k\neq j}P_{1}^{i}(k)T^{k}(j)\big)+(1-\alpha)b_{j},~~i,j,k=1,2,\cdots,N
\end{eqnarray*}
The matrix form is
\begin{eqnarray*}
P_{2}=\alpha P_{1}T+(1-\alpha)Y
\end{eqnarray*}
where
\begin{eqnarray*}
Y=\left(\begin{matrix}
b_{1}& b_{2}& \cdots& b_{N} \\
b_{1}& b_{2}& \cdots& b_{N} \\
\vdots& \vdots& \vdots& \vdots \\
b_{1}& b_{2}& \cdots& b_{N}
\end{matrix}
\right)
\end{eqnarray*}
Figure \ref{immigration} shows the relationship between the immigration mechanism and the branching-mutation mechanism. The left side of the figure represents the immigration mechanism, while the right side represents branching - mutation mechanism. The convergence of this process is given below.
\begin{eqnarray*}
F_{0}=P, T=P
\end{eqnarray*}
\begin{eqnarray*}
F_{n}=\alpha F_{n-1}T+(1-\alpha)Y=\alpha^{n}P^{n+1}+(1-\alpha)Y\Sigma_{k=0}^{n-1}\alpha^{k}P^{k}
\end{eqnarray*}
Since $T=P$ is regular which is explained in section 3.2 that meets the condition of Lemma \ref{convergence}, the limit $\lim_{n\to \infty}P^{n+1}$ exists. And for $\Sigma_{k=0}^{n-1}\alpha^{k}P^{k}$, since the eigenvalue of $\alpha P$, $\rho(\alpha P)<1$, the limit of $\Sigma_{k=0}^{n-1}(\alpha P)^{k}$ exists when $n\to \infty$. Consequently, the $\lim_{n\to \infty}f_{n}$ exists.

\subsection{The Google PageRank system }
There are optimal iteration algorithms for the diffusion process, one of the most successful methods is the Google PageRank system. The system was originally designed to objectively rank webpages by measuring people's interest in the relevant webpages and taking into account the potential hyperlink structure. In the Google PageRank system, in addition to walking on neighboring nodes with a probability $\alpha\in(0,1)$, the particle is also considered to jump to an arbitrary node with a very small probability $(1-\alpha)$, following the update mechanism below:
\begin{eqnarray*}
F_{n+1}=\alpha F_{n}P+(1-\alpha)Y
\end{eqnarray*}
$N\times N$ matrix $Y$ is a stacking of row vectors $\mathbf{y}$, and $\mathbf{y}$ defines the probabilities of randomly jumping to the corresponding nodes.

We use the MGWF with mutation and immigration model to explain the extension of the diffusion process. For convenience, we make the following transformation :
 \begin{eqnarray*}
F_{n+1}=\alpha F_{n}P+(1-\alpha)Y\\
=\left(\begin{matrix}
\alpha F_{n}&(1-\alpha) I
\end{matrix}
\right)\left(\begin{matrix}
P\\
Y
\end{matrix}
\right)
\end{eqnarray*}

In the above model, the branching law of the $n$-th generation is $\left(\begin{matrix}\alpha F_{n}&(1-\alpha) I\end{matrix}\right)$. Where the branching - mutation mechanism works with probability $\alpha$ and the branching law matrix is $F_{n}$. The immigration mechanism works with probability $(1-\alpha)$ and the branching law matrix is $I$; The probability matrix of mutation is $\left(\begin{matrix}P\\Y\end{matrix}\right)$ in mutation phase, where $P,~Y$ respectively are the mutation probability of MGWF with mutation mechanism and immigration mechanism.

We can explain the diffusion process and the Google PageRank system in the same framework model through the proposed two-phases setting. The initial branching law matrix is $F_{0}=\left(\begin{matrix}\alpha P&(1-\alpha) I\end{matrix}\right)$ in the branching phase. And the probability matrix of mutation is $T=\left(\begin{matrix}P\\Y\end{matrix}\right)$ in the mutation phase. The iteration rule is $F_{n+1}=F_{n} T$. When $\alpha=1$, it's diffusion process  $F_{n+1}=F_{n} P$; When $\alpha\in(0,1)$, it's the Google PageRank system $F_{n+1}=\alpha F_{n}P+(1-\alpha)Y$. As shown in Table \ref{framework}.

\begin{table*}[th]
\begin{center}
\caption{The same framework model of diffusion process and Google PageRank system }
\label{framework}
\begin{tabular}{lccc}
\hline
The framework&$\alpha$ &The model\\
\hline
$F_{n+1}=\left(\begin{matrix}\alpha F_{n}&(1-\alpha)I\end{matrix}\right) \left(\begin{matrix}P\\Y\end{matrix}\right)$&$\alpha=1$& The diffusion process\cr
{}&$\alpha\in(0,1)$& The Google PageRank system\cr

       \hline
\end{tabular}
\end{center}
\end{table*}

In summary, we proposed the two-phase MGWF setting and use it to explain the diffusion process and the Google PageRank system. This gives us a new perspective to understand the random walk. Further, we can use the MGWF framework to extend the diffusion process and speed up the iteration process as shown in the examples below.

\section{Simulations}
\label{Simulations}
\label{}
\begin{example}{The continuation of Example 3.2}\label{changeT}

We consider a diffusion process where the initial value $F_{0}=P$, the transition matrix $P$, and the iteration rule $F_{n+1}=F_{n}P$ with $P$ given below. We explain the diffusion process by random walk and MGWF respectively.
\begin{eqnarray*}
P=\left(\begin{matrix}
0& 0.6250&0.3750\\
0.3846&0&0.6154\\
0.2727&0.7273&0
\end{matrix}
\right).
\end{eqnarray*}
(i) Random walk on graph
\begin{figure}[H]
\centering
\includegraphics[width=1.0\textwidth]{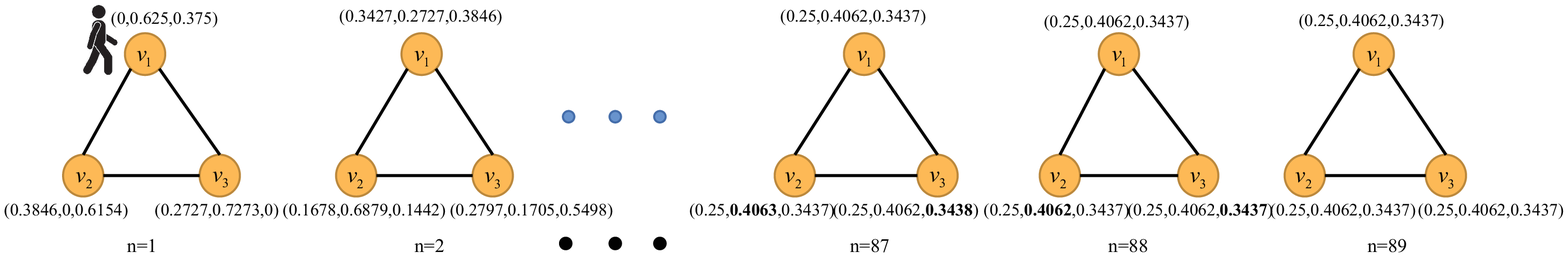}
\caption{Random walk on graph with 3 nodes}
\label{randomwalk2}
\end{figure}
For the initial value $F_{0}=P$, the probability that the particle starts at vertices 1,2, and 3 is $(\begin{matrix}0& 0.6250&0.3750\end{matrix}), (\begin{matrix}0.3846&0&0.6154\end{matrix})$ and $(\begin{matrix}0.2727&0.7273&0\end{matrix})$.
Although \cite{CL,LL} has verified that the selection of initial value is more efficient for convergence, the reason for such selection cannot be justified explicitly in the random walk. With the transition matrix $P$, and the iteration rule $F_{n+1}=F_{n}P$, we use the binary norm of the matrix $\|F_{n}-F_{n-1}\|_{2}$ to measure the convergence of $F_{n}$ with an accuracy of $1e^{-8}$. We find that the diffusion process can reach this approximate stable state through 54 iterations. The random walk on the graph with 3 nodes is shown in Figure \ref{randomwalk2}.

(ii) MGWF
\begin{figure}[H]
\centering
\includegraphics[width=0.6\textwidth]{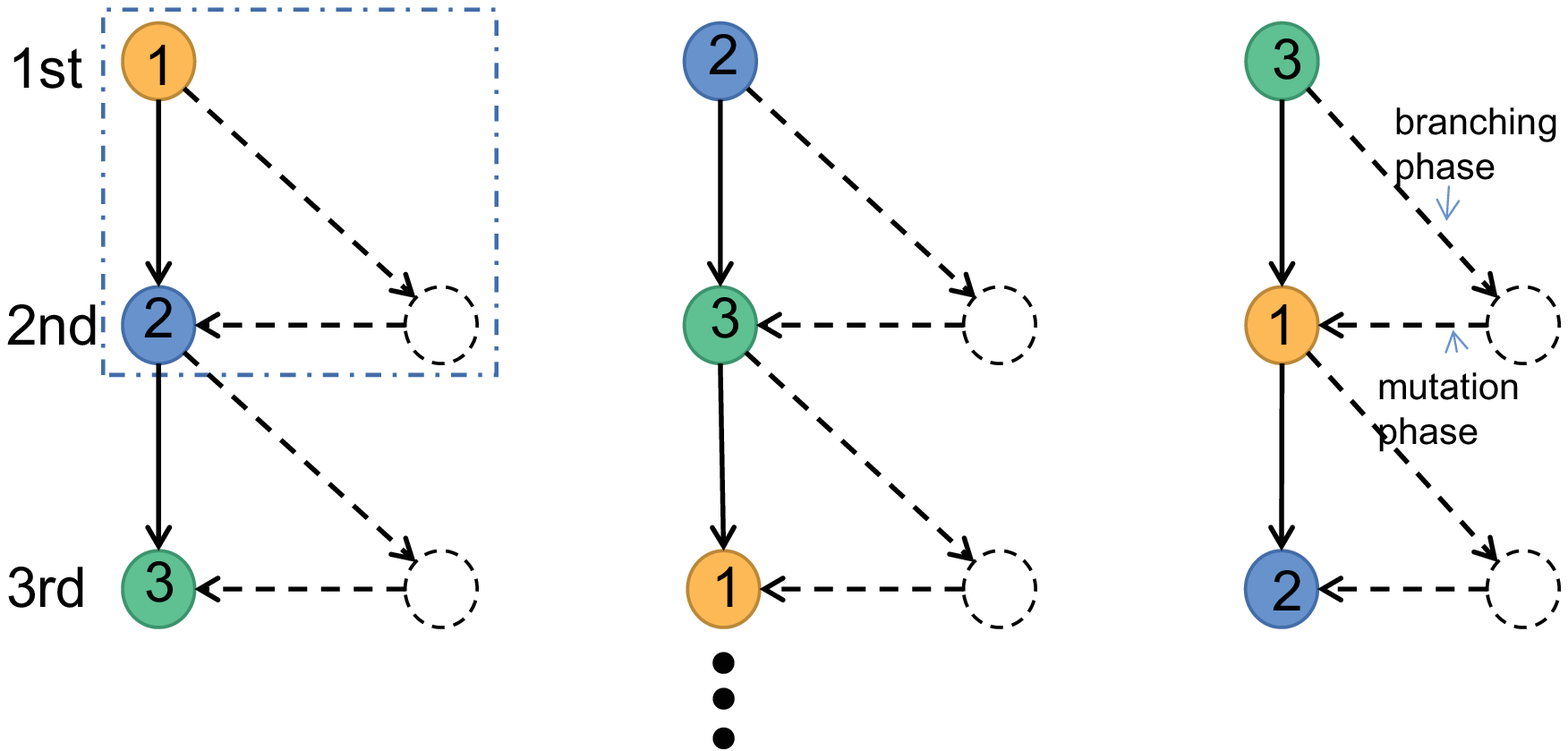}
\caption{MGWF with mutation}
\label{MGWF}
\end{figure}
In MGWF with mutation, the initial branching law is $F_{0}=P$, such as the probability of the parent of $1$-type producing children of $1, 2, 3$-type is $0, 0.625, 0.375$ respectively. From the figure above, it can be clearly observed that each unit time is divided into two phases in the $n$-th generation, where in the branching phase of the branching law is $F_n$, the probability matrix of mutation in the mutation phase is $T=P$. In order to get the branching law of $(n+1)$-th generation, we follow the result $F_{n+1}=F_{n}T$. It's shown in Figure \ref{MGWF}. It is also found that the approximate stable state with an accuracy of $1e^{-8}$ is basically reached after 54 iterations through simulation.

From the above description, we find that the model of MGWF with the mutation can intuitively explain the diffusion process better. Moreover, we can make the branching law matrix be the same in the branching phase, but the probability matrix of mutation is different in the odd iteration and even iteration mutation phase. In the odd iteration, the probability matrix of mutation is $T$, in the even iteration, we change the probability matrix of mutation to $\lambda T+(1-\lambda)I$ which is different from $T$ with $\lambda\in(0,1)$ and $I$ is an identity matrix. Such a slight change leads to a different diffusion process:
\begin{eqnarray*}
F_{n+1}=\begin{cases}F_{n} T,& \text{if} ~~n~~\text{is odd};\\ F_{n}(\lambda T+(1-\lambda)I),& \text{if} ~~n~~ \text{is even}.\end{cases}
\end{eqnarray*}

Let us take $\lambda=\frac{1}{3}$ and apply the revised diffusion process to the example of N=3 above. We also use the two-norm of the matrix $\|F_{n}-F_{n-1}\|_{2}$ to measure the convergence of $F_{n}$ with an accuracy of $1e^{-8}$. We found through the simulation that the diffusion process reached an approximate stable state after 17 iterations, which was significantly more efficient than the original diffusion process.

The convergence behaviors of the original diffusion process and the revised diffusion process is shown in Figure \ref{Iteration}.
\begin{figure}[H]
\centering
\includegraphics[width=0.6\textwidth]{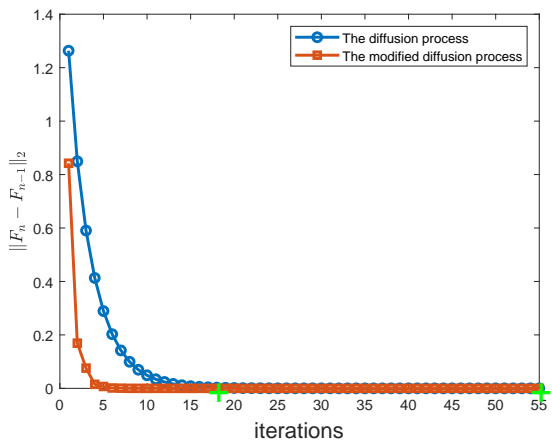}
\caption{ The convergence behaviors of the diffusion and the modify diffusion process}
\label{Iteration}
\end{figure}
\end{example}

\begin{example}\label{changeY}
We consider a Google PageRank system with the initial value $F_{0}=P$, the transition matrix $P$, the probability matrix of random jump $Y$, and the iteration rule $F_{n+1}=\alpha F_{n}P+(1-\alpha)Y$. We explain the Google PageRank system by MGWF with mutation and immigration with $\alpha=0.9$,
\begin{eqnarray*}
P=\left(\begin{matrix}
0& 0.6250&0.3750\\
0.3846&0&0.6154\\
0.2727&0.7273&0
\end{matrix}
\right).
\end{eqnarray*}
\begin{eqnarray*}
Y=\left(\begin{matrix}
0&0.4258&0.5742\\
0.2514&0&0.7486\\
0.4166&0.5834&0
\end{matrix}
\right).
\end{eqnarray*}

We use the two-phase theory in MGWF with mutation and immigration to explain the Google PageRank system. The initial branching law is $\left(\begin{matrix}\alpha P&(1-\alpha) I\end{matrix}\right)$. The probability matrix of mutation is $\left(\begin{matrix}P\\Y\end{matrix}\right)$ in the mutation phase. In order to get the branching law of $(n+1)$-th generation, we follow the result $\left(\begin{matrix}\alpha F_{n}&(1-\alpha) I\end{matrix}\right)\left(\begin{matrix}P\\Y\end{matrix}\right)$. We use the norm  $\|F_{n}-F_{n-1}\|_{2}$ to measure the convergence of $F_{n}$ with an accuracy of $1e^{-8}$. We find that the Google PageRank system can reach this approximate stable state through 42 iterations.

Whereas the random walk can't explain the Google PageRank system, the Google PageRank system is perfectly explained by considering the immigration in the MGWF with mutation. Similar to the basic diffusion process, we propose a modified PageRank system by dividing each unit of time into two phases:

(i)As in Example 5.1, we also made some minor changes to the model of the Google PageRank system. We make the branching law matrix is the same in the branching phase, and the probability matrix of mutation different in the odd iteration and even iteration mutation phase. In the odd iteration, the probability matrix of mutation is $\left(\begin{matrix}P\\Y\end{matrix}\right)$, in the even iteration, we change the probability matrix of mutation to $\left[\lambda\left(\begin{matrix}P\\Y\end{matrix}\right)+(1-\lambda)\left(\begin{matrix}I\\J\end{matrix}\right)\right]$ :

\begin{eqnarray*}
F_{n+1}=\begin{cases}\left(\begin{matrix}\alpha F_{n}&(1-\alpha) I\end{matrix}\right)\left(\begin{matrix}P\\Y\end{matrix}\right),& \text{if} ~~n~~\text{is odd};\\ \left(\begin{matrix}\alpha F_{n}&(1-\alpha) I\end{matrix}\right)\left[\lambda\left(\begin{matrix}P\\Y\end{matrix}\right)+(1-\lambda)\left(\begin{matrix}I\\J\end{matrix}\right)\right],& \text{if} ~~n~~ \text{is even}.\end{cases}
\end{eqnarray*}
where $\lambda\in(0,1)$, $I$ is an identity matrix and \begin{eqnarray*}
J=\left(\begin{matrix}
0&0.9086&0.0914\\
0.8412&0&0.1588\\
0.5715&0.4285&0
\end{matrix}
\right).
\end{eqnarray*} is another probability matrix of mutation in immigration mechanism which is different from $Y$.

Let us take $\lambda=\frac{3}{5}$ and apply the revised Google PageRank system to the example with N=3 above. We found that the system has reached two different stable states in the 7th iteration and the 8th iteration respectively, which makes the norm $\|F_{n}-F_{n-1}\|_{2}$ a fixed constant. The speed of reaching stability of the revised Google PageRank was faster than that of the original Google PageRank system. We will conduct an in-depth study on the relationship and the choice between the two stable states in the future.

The convergence behaviors of the Google PageRank system and the revised Google PageRank system is shown in Figure \ref{IterationPT+Y}.
\begin{figure}[H]
\centering
\includegraphics[width=0.6\textwidth]{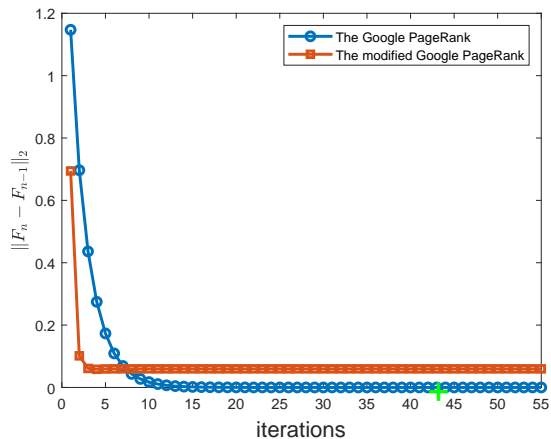}
\caption{ The convergence behaviors of the Google PageRank system and the revised  Google PageRank system}
\label{IterationPT+Y}
\end{figure}

(ii) We can make a couple of other changes to look at the branching-mutation mechanism and the immigration mechanism, whose change in mutation probability is responsible for the rate of convergence of $F_n$.

(a){$\mathbf{(The~change~in~the~branching-mutation~mechanism)}$} In the odd and even iterations, the different probability matrices of mutation in the branching-mutation mechanism are selected respectively and the probability matrices of mutation in the immigration mechanism are the same.

\begin{eqnarray*}
F_{n+1}=\begin{cases}\left(\begin{matrix}\alpha F_{n}&(1-\alpha) I\end{matrix}\right)\left(\begin{matrix}P\\Y\end{matrix}\right),& \text{if} ~~n~~\text{is odd};\\ \left(\begin{matrix}\alpha F_{n}&(1-\alpha) I\end{matrix}\right)\left(\begin{matrix}(\lambda P+(1-\lambda) I)\\Y\end{matrix}\right),& \text{if} ~~n~~ \text{is even}.\end{cases}
\end{eqnarray*}

We took $\lambda=\frac{2}{5}$ and found that the system would reach two different stable states in the 7th iteration and the 8th iteration respectively.

(b){$\mathbf{(The~change~in~the~immigration~mechanism)}$} In the odd and even iterations, different probability matrix of immigration $Y$ is selected respectively.

\begin{eqnarray*}
F_{n+1}=\begin{cases}\left(\begin{matrix}\alpha F_{n}&(1-\alpha) I\end{matrix}\right)\left(\begin{matrix}P\\Y\end{matrix}\right),& \text{if} ~~n~~\text{is odd};\\ \left(\begin{matrix}\alpha F_{n}&(1-\alpha) I\end{matrix}\right)\left(\begin{matrix}P\\\lambda Y+(1-\lambda) J\end{matrix}\right),& \text{if} ~~n~~ \text{is even}.\end{cases}
\end{eqnarray*}

We took $\lambda=\frac{2}{5}$ and found that the system reached two different stable states in the 11th iteration and the 12th iteration respectively.

According to the iteration results of (a) and (b) examples, it is found that changing the probability matrix of mutation in the branching-mutation mechanism is the key technique to improve the iteration speed. The convergence behaviors of (a),(b) are shown in Figure \ref{Iterationeg52}.
\begin{figure}[H]
\centering
\includegraphics[width=0.6\textwidth]{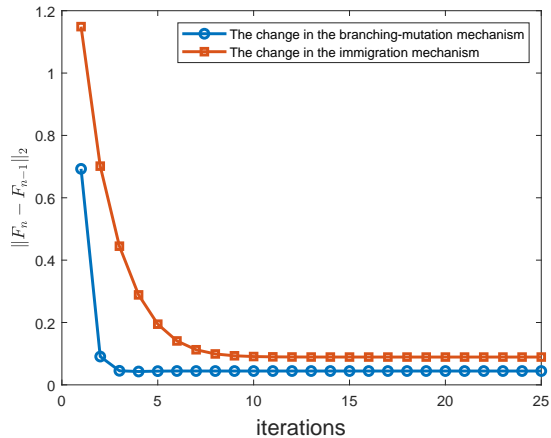}
\caption{ The convergence behaviors of the change of probability matrix of mutation in the branching mechanism  and in the immigration mechanism}
\label{Iterationeg52}
\end{figure}
\end{example}

\section{Conclusions}
\label{Conclusions}
In this paper, we have connected the two seemingly unrelated fields of diffusion process and branching process and established an explicit interpretation of the correspondence between the diffusion process and MGWF. Then we focused on MGWF with degenerated branching law, and innovatively proposed the concept of particle mutation in MGWF with degenerated branching law. Through the variant of MGWF, the diffusion process can be better interpreted, and we can observe the movement of particles from the longitudinal view more clearly. We divided each step of the diffusion process into two phases - branching phase and mutation phase, which not only interprets the mechanics more clear but also removes the existing limitation that the transition probability $P$ is dependent on the affinity matrix $A$. By extending MGWF with the concept of immigration, we further connected the MGWF model with the popular Google PageRank system, providing a new perspective to the latter as well.  Both the diffusion process and the Google PageRank are well explained with the two-phases idea, as shown in Table \ref{two phases}.

\begin{table*}[th]
\begin{center}
\caption{The diffusion process and the Google PageRank are explained with the two-phases idea }
\label{two phases}
\begin{tabular}{lcccc}
\hline
Method&{}&The branching law&{}&The probability of mutation\\
\hline
Diffusion process&{}&$F_{n}$&{}&$P$\cr
Google pagerank system&{}&$\left(\begin{matrix}\alpha F_{n}&(1-\alpha) I\end{matrix}\right)$&{}&$\left(\begin{matrix}P\\Y\end{matrix}\right)$\cr

       \hline
\end{tabular}
\end{center}
\end{table*}

In the future, we will focus on the following directions

\begin{itemize}
\item[$\bullet$] From Table \ref{two phases}, we found that the existing improvements of the diffusion process are concentrated in the branching phase, that is, to change the branching law by iteration. However, for the mutation phase, the mutation probability of all algorithms is constant. Therefore, in the future, we will pay more attention to changing the probability of mutation at each unit time.
\item[$\bullet$] In Example \ref{changeT}, we found that the rate of convergence of the diffusion process was significantly accelerated after we made a slight change in the mutation probability. We will study how to change the probability of mutation to accelerate the rate of convergence.
\item[$\bullet$] In Example \ref{changeY}, after slight changes were made to the mutation probability of the Google PageRank system, we found that although the rate of convergence of the Google PageRank system was significantly accelerated, two stable states appeared. It is of interest to study how to deal with two stable states.
\end{itemize}

\end{document}